\documentclass[final]{siamltex}

\usepackage{amsmath,amssymb,graphicx,graphics}
\usepackage[dvips]{epsfig}
\usepackage{showkeys}

\def \R{\mathbb R}

\newtheorem{remark}{Remark}[section]

\title{Multi-scale CLEAN in hard X-ray solar imaging}

\author{A. Volpara\thanks{MIDA, Dipartimento di Matematica, Universit\`a di Genova, via Dodecaneso 35, 16146 Genova, Italy  ({\tt volpara@dima.unige.it})} 
\and
M. Piana\thanks{MIDA, Dipartimento di Matematica, Universit\`a di Genova, via Dodecaneso 35, 16146 Genova, Italy \& Istituto Nazionale di Astrofisica, Osservatorio Astrofisico di Torino, Torino, Italy ({\tt piana@dima.unige.it})}
\and
A. M.~Massone\thanks{MIDA, Dipartimento di Matematica, Universit\`a di Genova, via Dodecaneso 35, 16146 Genova, Italy ({\tt massone@dima.unige.it}) }}

\begin{document}

\maketitle

\begin{abstract}
Multi-scale deconvolution is an ill-posed inverse problem in imaging, with applications ranging from microscopy, through medical imaging, to astronomical remote sensing. In the case of high-energy space telescopes, multi-scale deconvolution algorithms need to account for the peculiar property of native measurements, which are sparse samples of the Fourier transform of the incoming radiation. The present paper proposes a multi-scale version of CLEAN, which is the most popular iterative deconvolution method in Fourier space imaging. Using synthetic data generated according to a simulated but realistic source configuration, we show that this multi-scale version of CLEAN performs better than the original one in terms of accuracy, photometry, and regularization. Further, the application to a data set measured by the NASA Reuven Ramaty High Energy Solar Spectroscopic Imager (RHESSI) shows the ability of multi-scale CLEAN to reconstruct rather complex topographies, characteristic of a real flaring event.
\end{abstract}



\pagestyle{myheadings}

\thispagestyle{plain}

\markboth{A. Volpara et al}{multi-scale CLEAN}

\section{Introduction}
Both on earth and from space astronomical data require a notable pre-processing effort before they can be transformed into meaningful images and then utilized for physical interpretation. In fact, the main mathematical challenge of astronomical image reconstruction \cite{pantin2017deconvolution} is the need of solving an ill-posed inverse problem whose instability issues can be addressed by means of regularization constraints \cite{bertero2021introduction}. In particular, modalities like on-earth radio \cite{kruger2012introduction} and from-space hard X-ray imaging \cite{piana2022hard} work by temporal source modulation rather than focusing, so that these telescopes measure a specific set of spatial Fourier components of the source, called visibilities, sampled in correspondence of specific spatial frequency pairs, called $(u,v)$ points. Image reconstruction from visibilities is, therefore, a Fourier transform inversion problem from limited data based on two steps:
\begin{itemize}
\item The gridding step transforms the visibility signal recorded in the $(u,v)$ plane into a (blurred) image.
\item The deconvolution step subtracts the effect of the Point Spread Function (PSF) of the instrument.
\end{itemize}
This paper focuses on this second deconvolution step in the case of hard X-ray space data, and introduces a multi-scale formulation of the most used deconvolution algorithm in this framework, i.e., the CLEAN algorithm \cite{hogbom1974aperture,dennis2009hard,li2011application,zhang2020parameterized,chu2019deconvolution,bose2002sequence}. In particular, X-ray collimators perform a sparse and limited sampling in the $(u,v)$ plane. This implies that inverse Fourier transforming X-ray visibilities yields notable artifacts in the image provided by gridding. In the CLEAN framework this image is named dirty map and the technical goal of this algorithm is to non-linearly and iteratively deconvolve the instrumental PSF (named dirty beam in this context) from the dirty map by
\begin{enumerate}
\item Identifying the maximum and its position.
\item Generating a CLEAN component, i.e. a $\delta$-Dirac, located in this position in a twin image called  the CLEAN components map.
\item Subtracting the PSF centered on the position of the maximum from the dirty map.
\item Repeating this scheme until the remaining dirty map  contains just noise.
\item Generating the CLEANed image as the convolution of the collection of CLEAN components with an idealized PSF, called CLEAN beam, obtained by fitting the dirty beam by means of a 2D Gaussian function.
\item Adding appropriate residuals to the CLEANed image.
\end{enumerate}
This algorithm is fast and effective in removing the sidelobes introduced by the PSF. However, CLEAN introduces significant photometric errors (the total flux is not conserved during the iterations) and, more importantly, in complex images is prone to misinterpret the spatial scales concurrently characterizing the different sources. 

This paper introduces a multi-scale version of CLEAN for deconvolving hard X-ray data which is partly inspired by an algorithm introduced in 2008 in radio imaging \cite{cornwell2008multiscale}. The formulation here presented is specifically tailored on a specific property of all the hard X-ray instruments of most recent conception and, as remarked in Section 3, it differs from its radio-interferometry counterpart at both a conceptual and a technical level. In fact, {\em{YohKoh}} (operating over 1991-2001) \cite{1992Sci...258..618A,ogawara1992status}, the Reuven Ramaty High Energy Solar Spectroscopic Imager (RHESSI) (launched by NASA in 2002 and decommissioned in 2018) \cite{lin2003reuven,fletcher2011observational} and the Spectrometer/Telescope for Imaging X-rays (STIX) (which has begun its travel to the Sun in February 2020 in the ESA Solar Orbiter payload) \cite{krucker2020spectrometer,massa2022first}, all sample the $(u,v)$ plane according to circles characterized by increasing radii. This configuration implies 
that the PSF of these telescopes is the sum of PSF components, each one associated to a specific circle of sampled visibilities in the $(u,v)$ plane. Accordingly, the dirty map corresponding to the measured hard X-ray visibilities can be straightforwardly interpreted as the sum of several dirty maps, each one corresponding to one PSF component. Finally, this property has a technical impact on two implementation steps: the way the dirty maps are rescaled in the pre-processing step and the way residuals are added to the CLEANed image in the final step of the algorithm. 

This formulation of the CLEAN method is genuinely multi-scale, since it is able to point out all different scales using, in one shot, the whole dataset and the total PSF. Further, its implementation is naturally biased in favor of the reconstruction of smaller sources, that, otherwise, could be obscured by sources with larger scales. We showed the effectiveness of our approach by first validating the algorithm against synthetic but realistically simulated visibility sets. Then, we applied the method to an experimental data set observed by RHESSI during an intense X-ray flare.

The plan of the paper is as follows. In Section 2 the well-known CLEAN algorithm for visibilities is recalled. Section 3 introduces the new multi-scale CLEAN approach for hard X-ray visibilities. Section 4 provides the numerical validation of the method against synthetic visibilities mimicking RHESSI acquisition. Section 5 describes an application to experimental RHESSI data. Our conclusions are offered in Section 6.

\section{The CLEAN algorithm: mathematical setup}
In radio and X-ray astronomical imaging, CLEAN is a numerical procedure for the solution of the convolution equation
\begin{equation}\label{prob}
I^D(x,y) = (K * I)(x,y) := \int\int K(x-x',y-y')I(x',y')dx'dy'\ \ ,
\end{equation}
where $I$ is the unknown source flux, $I^D$, the dirty map, is the input blurred and noisy image, and $K$ is the Point Spread Function (PSF) of the instrument. CLEAN models the solution $I(x,y)$ as the sum of $Q$ unknown point sources described by $\delta$-Dirac distributions plus background, i.e.:
\begin{equation}\label{prob-1}
I(x,y) = \sum_{q=1}^Q I_q \delta(x-x_q,y-y_q) + B(x,y) \ ,
\end{equation}
where $I_q$ and $(x_q,y_q)$ are respectively the amplitude and position of the $q$-th source and $B(x,y)$ models the background. Using (\ref{prob-1}) into (\ref{prob}) leads to
\begin{equation}\label{dirty}
I^D(x,y) 
         = \sum_{q=1}^Q I_q K(x-x_q,y-y_q) + (K*B)(x,y).
\end{equation}

The goal of CLEAN is then to iteratively estimate $I_q, x_q, y_q$ for $q=1,\ldots,Q$, as well as $Q$, given an estimate of the dirty map $I^D$ and armed with the instrument PSF $K$, here called dirty beam. Starting from the initialization
\begin{equation}\label{step-0}
I^{(0)}(x,y) := I^D(x,y) 
\end{equation}
the following CLEAN loop is at the base of such algorithm:
\vspace{0.25cm}

\begin{description}
\item{{\bf{Maximum Identification.}}} This is an easy optimization step that provides, at the $t$-th iteration of the loop $(t\ge1)$, the maximum and its position in the dirty image by computing
\begin{equation}\label{step-1}
(x_{max}^{(t)},y^{(t)}_{max}) = \arg\max_{(x,y)} I^{(t-1)}(x,y)~~~~~~~~~~~I^{(t)}_{max} = I^{(t-1)}(x_{max}^{(t)},y_{max}^{(t)}).
\end{equation}
\item{{\bf{CLEAN Components Update.}}} Starting from an empty map $CC^{(0)} (x,y)$, the {\it CLEAN components map} at the $t$-th iteration is
\begin{equation}\label{step-2}
CC^{(t)} (x,y) = CC^{(t-1)}(x,y) + \frac{\gamma I^{(t)}_{max}}{\max_{(x,y)} |K(x,y)|} \delta(x-x_{max}^{(t)},y-y_{max}^{(t)}),
\end{equation}
where $\gamma$ is the so-called {\em{gain factor}}.
\vspace{0.2cm}
\item{{\bf{Dirty Map Update.}}} The dirty beam centered at the maximum is subtracted from the actual dirty map (i.e., $I^{(t-1)}(x,y)$) to obtain
\begin{equation}\label{step-3}
I^{(t)}(x,y) = I^{(t-1)}(x,y) - \frac{\gamma I^{(t)}_{max}}{\max_{(x,y)} |K(x,y)|} K(x-x_{max}^{(t)},y-y_{max}^{(t)}).
\end{equation}
\item{{\bf{Update.}}} $t=t+1$ and then go back to the first step until a stopping rule is verified.
\end{description}

\vspace{0.25 cm}

The idea behind the CLEAN loop is the one to iteratively identify point sources in the actual dirty map and to save their intensity (scaled by the gain factor and the PSF peak) and position in the CLEAN components map. At the same time, in the Dirty Map Update step, the blurring effect of the PSF on the same point source is removed from the dirty map by subtracting the (rescaled) dirty beam.

The stopping rule for the loop applies when there are no more point sources in the updated dirty map, which occurs when this is embedded by the background map (i.e., when just positive and negative lobes of the same amplitude survive in the updated dirty map). Therefore, if $t={\overline{t}}$ is the iteration when this happens, then we define the residual map
\begin{equation}\label{residual-map}
R(x,y) := I^{({\overline{t}})}(x,y)
\end{equation}
and, from eq.~(\ref{dirty}), we have 
\begin{equation}\label{final-residual}
R(x,y) \simeq (K * B)(x,y),
\end{equation}
which inspires the following approximation for the background map:
\begin{equation}\label{background}
B(x,y) \simeq B^{\prime}(x,y):= \frac{R(x,y)}{T},
\end{equation}
where $T$ is an estimate of the integral of the PSF over the Field of View.

As the final step of CLEAN deconvolution, one constructs the CLEANed map $C(x,y)$ as 
\begin{equation}\label{CLEAN-final}
C(x,y) = (K^C * (CC^{({\overline{t}})}+B^{\prime}))(x,y)=(K^C * CC^{({\overline{t}})})(x,y) +  (K^C * B^{\prime})(x,y),
\end{equation}
where $K^C(x,y)$ is an idealized version of the instrumental, experimental PSF, here called CLEAN beam.

\vspace{0.25 cm}
\begin{remark}
The gain factor controls the fraction of each point source to be added in the CLEAN components map and subtracted from the dirty map. Typical values are in the range $[0.05,0.5]$.
\end{remark}
\vspace{0.25 cm}
\begin{remark} An estimate for $T$ is given by $T = \sum_{n}^{N_{pix}} K(x_n,y_n) \delta x \delta y$, where $(x_n,y_n)$ is the $n$-th pixel ($n=1,\dots,N_{pix}$) in the PSF image and $\delta x$, $\delta y$ are the pixel dimensions. 
\end{remark}

\subsection{The CLEAN algorithm in hard X-ray imaging}
In hard X-ray (and in radio astronomy) imaging, the experimental observations are given as a set of $N_V$ measured visibilities $\{V(u_l,v_l)\}_{l=1}^{N_V}$. Visibilities are samples of the Fourier transform of the source flux, i.e.
\begin{equation}\label{visibilities}
V(u_l,v_l) = \int_{\R^2} I(x,y) \exp[2 \pi i (u_l x  + v_l y)] dx dy~~~~l=1,\ldots,N_V.
\end{equation}
It follows that a straightforward estimate of the dirty map can be obtained by numerically computing the inverse Fourier transform of the set of sampled visibilities, i.e. 
\begin{equation}\label{dirty-1}
I^D(x,y) =  \sum_{l=1}^{N_V} V(u_l,v_l) \exp[-2 \pi i (x u_l + y v_l)] \delta u_l \delta v_l \ ,
\end{equation}
where $\delta u_l, \delta v_l$ are the weights in the numerical integration. 
From (\ref{visibilities}) it follows that $I(x,y)=\delta(x,y)$ implies $V(x,y)=1$ for all sampled spatial frequencies. Therefore, the dirty beam can be computed as
\begin{equation}\label{dirty-beam}
K(x,y) =  \sum_{l=1}^{N_V}  \exp[-2 \pi i (x u_l + y v_l)] \delta u_l \delta v_l \ ,
\end{equation}
i.e., the dirty beam is the discretized inverse Fourier transform of the characteristic function of the sampled visibilities (i.e, the transfer function of the imaging system). 
Finally, the idealized version of the PSF, the CLEAN beam, is obtained by fitting $K(x,y)$ in (\ref{dirty-beam}) with a two-dimensional Gaussian function.


\section{The multi-scale CLEAN algorithm}
The differences between CLEAN as presented in the previous section and our multi-scale version of the method are at a both model and algorithm level. 

\begin{figure}[h]
\begin{center}
\includegraphics[width=11.cm]{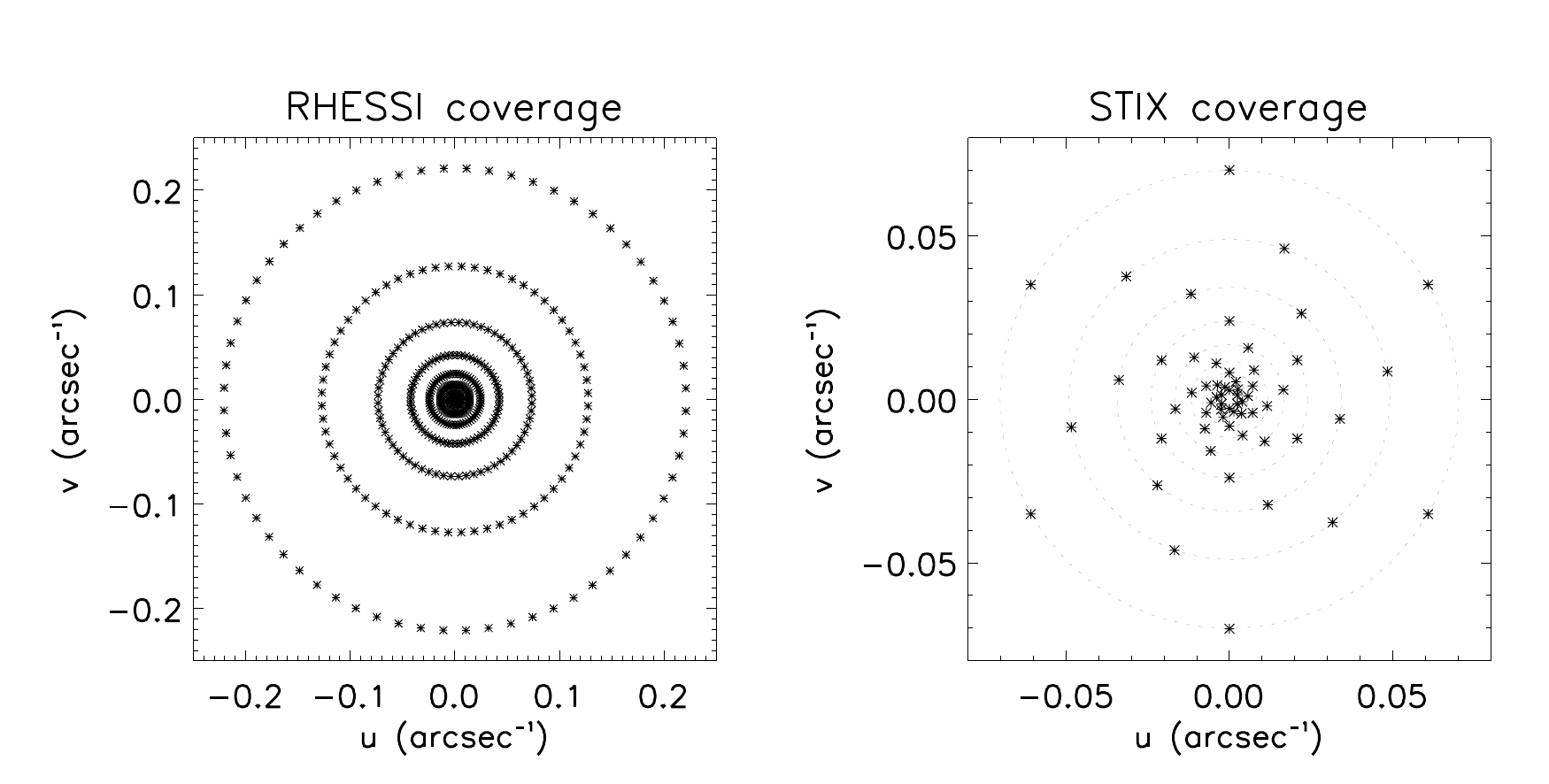} 
\caption{Sampling of the $(u,v)$ plane as performed by three hard X-ray telescopes. Left: {\em{RHESSI}}. Right: {\em{STIX}}.}
\label{fig:uv-samples}
\end{center}
\end{figure}

From the modeling viewpoint, we exploit the fact that hard X-ray telescopes are characterized by a PSF that can be written, under some assumptions, as the sum of a finite number of PSF components, each one filtering a specific portion of the $(u,v)$ plane.  For example, 
the NASA satellite RHESSI measured visibilities by sampling them over nine circles in the $(u,v)$ plane with radii decreasing from $R_1 \sim 0.22$ arcsec$^{-1}$ to $R_9 \sim 0.0027$ arcsec$^{-1}$, according to a geometric progression with ratio $\frac{1}{\sqrt{3}}$ (see Figure \ref{fig:uv-samples}, left panel) \cite{hurford2003rhessi}, where the number of samples in each circle is not fixed but determined in an optimal way during the computational procedure of visibility generation (data stacking). The ESA instrument STIX samples the $(u,v)$ plane in $60$ points\footnote{STIX recording hardware is made of 30 sub-collimators, each one sampling the spatial frequency domain in one specific point. However, the property $V(-u,-v) = \overline{V(u,v)}$ allows the duplication of the 30 spatial frequency samples, where the overline indicates the complex conjugate.} placed over six spirals, but in such a way that these points lay over $10$ circles with radii decreasing from $\sim 0.07$ arcsec$^{-1}$ to $\sim 0.0028 $ arcsec$^{-1}$, according to a geometric progression with ratio $\sim 0.7$ (see Figure \ref{fig:uv-samples}, right panel) \cite{massa2023stix} (more than thirty years ago also the Yohkoh Japanese mission picked-up Fourier components over six semi-circles with uniformly increasing radii \cite{kosugi1991hard}). Therefore, if we group the circles in the $(u,v)$ plane in $N$ disjoint sets, we can define a number $N$ of experimental PSF components, i.e. of dirty beams $\{K_j(x,y)\}_{j=1}^N$, such that
\begin{equation}\label{psf-component}
K_j(x,y) = \sum_{l=1}^{N_j} \exp[-2 \pi i (x u^{(j)}_l + y v^{(j)}_l)] \delta u^{(j)}_l \delta v^{(j)}_l~~~~~j=1,\ldots,N,
\end{equation}
where $N_j$ and $\{u_l^{(j)},v_l^{(j)}\}_{l=1}^{N_j}$ are the number and the set of visibilities belonging to the $j$-th subset of sampled circles in the $(u,v)$ plane, respectively. For example, in RHESSI we may have maximum $N=9$ dirty beams, which happens when each subset is made of just one of the $9$ circles sampled by the RHESSI Rotation Modulation Collimators. Similarly, in STIX we may have maximum $N=10$ dirty beams. Further,  under the assumptions that $\sum_{j=1}^N N_j = N_V$, the sets $\{u_l^{(1)},v_l^{(1)}\}_{l=1}^{N_1},\ldots,\{u_l^{(N)},v_l^{(N)}\}_{l=1}^{N_N}$ are all disjoint, and that natural weighting is considered (i.e. $\delta u^{(j)}_l=\delta u$ and $\delta v^{(j)}_l=\delta v$ for each $j=1,\dots,N$ and $l=1,\dots,N_j$), 
then 
\begin{equation}\label{psf-sum}
K(x,y) = \sum_{j=1}^N K_j (x,y)
\end{equation} 
where the $j$-th dirty beam reads
\begin{equation}\label{psf-natural}
K_j(x,y) = \sum_{l=1}^{N_j} \exp[-2 \pi i (x u^{(j)}_l + y v^{(j)}_l)] \delta u\delta v~~~~~j=1,\ldots,N.
\end{equation}
\begin{remark}
In the following we will assume that $K_1(x,y)$ is the highest resolution PSF component, i.e. that the corresponding transfer function samples the portion of the $(u,v)$ plane characterized by the highest frequencies. We will also assume that this resolution decreases as $j$ increases.  
\end{remark}
\vspace{0.25 cm}

Using the same approach, we can define a set of $N$ dirty maps
\begin{equation}\label{dirty-multi-1}
I^D_j(x,y) =  \sum_{l=1}^{N_j} V(u^{(j)}_l,v^{(j)}_l) \exp[-2 \pi i (x u^{(j)}_l + y v^{(j)}_l)] \delta u \delta v~~~~~j=1,\ldots,N
\end{equation}
such that 
\begin{equation}\label{dirty-sum}
I^D(x,y) = \sum_{j=1}^N I^D_j (x,y).
\end{equation}

Coherently with the model \eqref{psf-sum} for the PSF, the new model for the source image reads as the sum of $N$ basis functions $\{m_i\}_{i=1}^N$ each one characterized by a specific scale (for example, the support in the case of paraboloids, the Full Width at Half Maximum or the standard deviation in the case of Gaussian functions). 

\vspace{0.25 cm}
\begin{remark}
The geometrical scales $\{m_i\}_{i=1}^N$ represent the different physical scales showing up in the radiation source. Accordingly to Remark 3.1, we assume that $m_1(x,y)$ corresponds to the smallest scale and that the scale dimension increases as $i$ increases.
\end{remark}
\vspace{0.25 cm}

The source image $I(x,y)$ can be modeled as the superposition of the basis functions, i.e. as
\begin{equation}\label{source-multi}
I(x,y) = \sum_{i=1}^N\sum_{q_i=1}^{Q_i}I_{q_i}m_i(x-x_{q_i},y-y_{q_i}) + B(x,y),
\end{equation}
which means that at scale $i$, for $i=1,\ldots,N$, there are $Q_i$ sources, each one placed at $(x_{q_i},y_{q_i})$ and with peak intensity $I_{q_i}$ for $q_i = 1,\dots,Q_i$ (we assume here that each basis function $m_i(x,y)$ is normalized in such a way that $\int_{\R^2} m_i(x,y) dx dy=1$). Inserting equation (\ref{source-multi}) into the model equation (\ref{prob}) leads to
\begin{equation}
I^D(x,y) 
= \sum_{i=1}^N \sum_{q_i=1}^{Q_i} I_{q_i}  (m_i*K)(x-x_{q_i},y-y_{q_i}) + (K*B)(x,y).
\end{equation}
We now exploit equations (\ref{psf-sum}) and (\ref{dirty-sum}) to obtain

\begin{equation}\label{quasi-final2}
 \sum_j^N I_j^D(x,y) = \sum_{j=1}^N \left( \sum_{i=1}^N \sum_{q_i=1}^{Q_i} I_{q_i}  (m_i * K_j)(x-x_{q_i},y-y_{q_i}) + (K_j*B)(x,y) \right)  
\end{equation}

This equation inspires our multi-scale version of the CLEAN algorithm in which each dirty map $I_j^D(x,y)$ is identified as
\begin{equation}\label{final}
I_j^D(x,y) = \sum_{i=1}^N \sum_{q_i=1}^{Q_i} I_{q_i}  (m_i*K_j)(x-x_{q_i},y-y_{q_i}) + (K_j*B)(x,y),
\end{equation}
and CLEANed according to a parallel-wise procedure. The key idea is based on the interplaying properties of the basis functions and the PSF components. In fact,  equation (\ref{final}) shows that each dirty map contains cross-convolution products between the corresponding PSF component and all the basis functions. 

\vspace{0.25 cm}
\begin{remark}
Computing these cross-convolution products in the $(u,v)$ plane one immediately notes that  $m_i * K_j$ with $i>j$ are negligible. In fact, let us consider the Fourier Transforms $\widehat{m}_i(u,v)$ and $\widehat{K}_j(u,v)$ of $m_i(x,y)$ and $K_j(x,y)$ respectively; since the support of $\widehat{K}_j(u,v)$ is given by the set $\{u_l^{(j)},v_l^{(j)}\}_{l=1}^{N_j}$ of points sampled by the $j$-th set of detectors then if $i>j$ (i.e., the resolution of the basis function is lower then the resolution of the PSF) such points are outside the support of $\widehat{m}_i(u,v)$ and  $\widehat{m}_i(u,v)\cdot\widehat{K}_j(u,v)=0$. 
\end{remark}
\vspace{0.25 cm}

The strategy is therefore to identify small scale sources first and then to subtract their contribution to the whole set of dirty maps, provided that the identification of smaller scales requires a pre-processing step in which a bias towards these scales is applied.

Starting from the initialization  
\begin{equation}\label{step-0-multi}
I_j^{(0)}(x,y) := I_j^D(x,y) ~~~~~~~~~~
M_j:=\max_{(x,y)} (I_j^{(0)}(x,y))
\end{equation}
where $j=1,\ldots,N$ and estimates for $I_j^D(x,y)$ can be obtained by using equation (\ref{dirty-multi-1}), the multi-scale CLEAN algorithm can be summarized as the following loop: 
\vspace{0.25cm}

\begin{description}
\item{{\bf{Rescaling.}}} At each iteration $t\ge 1$, all dirty maps are re-scaled in such a way that the re-scaled dirty map components 
${\mathcal{I}}_j^{(t)}$, $j=1,\ldots,N$ become 
\begin{equation}\label{rescaling}
{\mathcal{I}}_j^{(t)}(x,y) = \eta_j \frac{I_j^{(t-1)}(x,y)}{M_j}
\end{equation}
where $\eta_j$, $j=1,\ldots,N$ are scale bias factors decreasing as $j$ increases. In this way, small scales will be favored at the first iterations. This rescaling is just a technical fact and re-scaled maps are utilized just in the maximum search process.
\vspace{0.2cm}
\item{{\bf{Scale Identification}}} In this step we determine the ${\overline{j}}$-th re-scaled dirty map that presents the maximum intensity value (at $t=1$ we will have ${\overline{j}}=1$)
\begin{equation}\label{multi-step-0}
\overline{j} = \arg\max_{j} {\mathcal{I}}_j^{(t)}(x,y),
\end{equation}
and we select the corresponding basis function $m_{\overline{j}}(x,y)$. 
\vspace{0.2cm}
\item{{\bf{Maximum Identification.}}} 
This optimization step provides the maximum and its position in the ${\overline{j}}$-th dirty image, i.e. it computes
\begin{equation}\label{multi-step-1}
(x_{max}^{(t)},y^{(t)}_{max}) = \arg\max_{(x,y)} I_{\overline{j}}^{(t-1)}(x,y)~~~~~~~~~~~I^{(t)}_{max} = I_{\overline{j}}^{(t-1)}(x_{max}^{(t)},y_{max}^{(t)}).
\end{equation}

\item{{\bf{CLEAN Components Update.}}} 
The multi-scale CLEAN components map at the $t$-th iteration is
\begin{equation}\label{multi-step-2}
CC^{(t)} (x,y) = CC^{(t-1)}(x,y) + \frac{ \gamma I^{(t)}_{max}}  {\max\limits_{ (x,y) }   |(m_{\overline{j}} * K_{\overline{j}})(x,y)|}m_{\overline{j}}(x-x_{max}^{(t)},y-y_{max}^{(t)}),
\end{equation}
where, again, $CC^{(0)}$ has all elements equal to zero and $\gamma$ is the {\em{gain factor}}.
\vspace{0.2cm}

\item{{\bf{Dirty Map Update.}}} At each scale $j=1,\ldots,N$, the dirty map component is then updated in such a way that
\begin{equation}\label{multi-step-3}
I_j^{(t)}(x,y) = I_j^{(t-1)}(x,y) - \frac{\gamma I^{(t)}_{max}}{\max\limits_{(x,y)} |(m_{\overline{j}} * K_{\overline{j}})(x,y)|} (m_{\overline{j}}*K_j)(x-x_{max}^{(t)},y-y_{max}^{(t)}).
\end{equation}
\item{{\bf{Update.}}} $t=t+1$ and then go back to the first step until a stopping rule is verified.

\end{description}
\vspace{0.2cm}

The stopping rule for the loop applies when there are no more sources in all updated dirty map components, which occurs when these maps are embedded by the corresponding background maps. 
Therefore, if $t={\overline{t}}$ is the iteration when this happens, then the residual map is defined as
\begin{equation}\label{multi-final-residual}
R(x,y) := \sum_{j=1}^N I_j^{(\overline{t})}(x,y) \simeq \sum_{j=1}^N (K_j *B)(x,y) =(K*B)(x,y),
\end{equation}
which inspires the usual approximation for the background map
\begin{equation}\label{multi-background}
B(x,y) \simeq B^{\prime}(x,y):= \frac{R(x,y)}{T},
\end{equation}
where, as in the CLEAN algorithm, $T$ is an estimate of the integral of the full PSF over the Field of View, i.e. $T = \sum_{n}^{N_{pix}} K(x_n,y_n) \delta x \delta y$, being $(x_n,y_n)$ the $n$-th pixel ($n=1,\dots,N_{pix}$) in the full PSF image and $\delta x$, $\delta y$ the pixel dimensions.

In the final step of multi-scale CLEAN algorithm, no convolution with any CLEAN beam is needed and one simply constructs the CLEANed map $C(x,y)$ as 
\begin{equation}\label{multiCLEAN-final}
C(x,y) = CC^{({\overline{t}})}(x,y)+B^{\prime}(x,y).
\end{equation}

\vspace{0.25 cm}
\begin{remark}
The definition of the scale bias $\eta_j$ is crucial in order to emphasize smaller scales at first iterations. In \cite{cornwell2008multiscale}, the author suggests a linear relation assigning decreasing weights to each scale as they increase. Here we use a relation reflecting the geometric progression of the radii of the circles sampling the $(u,v)$ plane. In particular, if $r$ is the ratio of the geometric progression ($r=\frac{1}{\sqrt{3}}$ for RHESSI) 
and the $(u,v)$ points are
grouped in $N$ sets in such way that the $j$-th set contains the $N_j$ sampling points laying on $C_j$ circles whose radii sorted in decreasing order are $\{ R_{h^j_1}, R_{h^j_2}, \dots , R_{h^j_{C_j}}\} \subset \{ R_1, R_2, ... , R_C \} $ ($C=9$ for RHESSI), 
we have defined:
\begin{equation}\label{eq:eta}
\eta_j:= r^{h_1^{j}-1} ~~~~~~~~ j=1,\dots,N.
\end{equation}
In the RHESSI case, for example, the sampled visibilities can be grouped in $N=3$ sets, where the first set may contain visibilities belonging to $C_1=3$ circles, the second one may contain visibilities from $C_2=3$ circles and the last one from $C_3=3$ circles. This means that the radii of the circles drawn by the visibilities of the first set are $\{ R_1, R_2, R_3 \}$ (i.e., $h^1_1=1$,  $h^1_2=2$, $h^1_3=3$), the ones of the second set are  $\{ R_4, R_5,R_6 \}$ (i.e., $h^2_1=4$, $h^2_2=5$ and $h^2_3=6$) and finally the ones of the last set are $\{ R_7,R_8,R_9 \}$ (i.e., $h^3_1=7$, $h^3_2=8$, and $h^3_3=9$). According to (\ref{eq:eta}), just  $h^1_1$,  $h^2_1$ and $h^3_1$ are used for the computation of the three scale bias $\eta_j$, $j=1,2,3$, leading to $\eta_1=1$, $\eta_2=0.19$ and $\eta_3=0.04$.
\end{remark}

\vspace{0.25 cm}
\begin{remark}
As in the CLEAN algorithm, the gain factor controls the fraction of each point source to be added in the clean components map and subtracted from the dirty maps. Typical values are in the range $[0.05,0.5]$.
\end{remark}
\vspace{0.25 cm}

\vspace{0.25 cm}
\begin{remark}
This multi-scale version of CLEAN differs from the one introduced in \cite{cornwell2008multiscale} for radio-imaging contexts at both conceptual and technical levels. Conceptually, here we fully exploit the fact that the PSF of hard X-ray telescopes can be decomposed as the sum of several components. Each one of these components can be used to compute the cross-convolution product with the basis functions, without the need to approximate it as the convolution between each basis function and the global PSF as done in \cite{cornwell2008multiscale}. From a technical viewpoint, here we use two-dimensional Gaussian functions (and not two-dimensional paraboloids) as basis functions, which is better appropriate to the fact that in hard X-ray imaging the flaring sources do not have a compact support. Further, the scale bias is computed in a completely different way and, also, our scheme does not involve a final convolution step with the CLEAN beam as in the Cornwell's algorithm. Finally, we remark that this is the first time that a multi-scale CLEAN algorithm is applied to hard X-ray visibilities using both synthetic and experimental data.
\end{remark}
\vspace{0.25 cm}

\section{Applications}
\begin{figure}[h]
\begin{center}
\includegraphics[width=7.6cm, angle=90]
{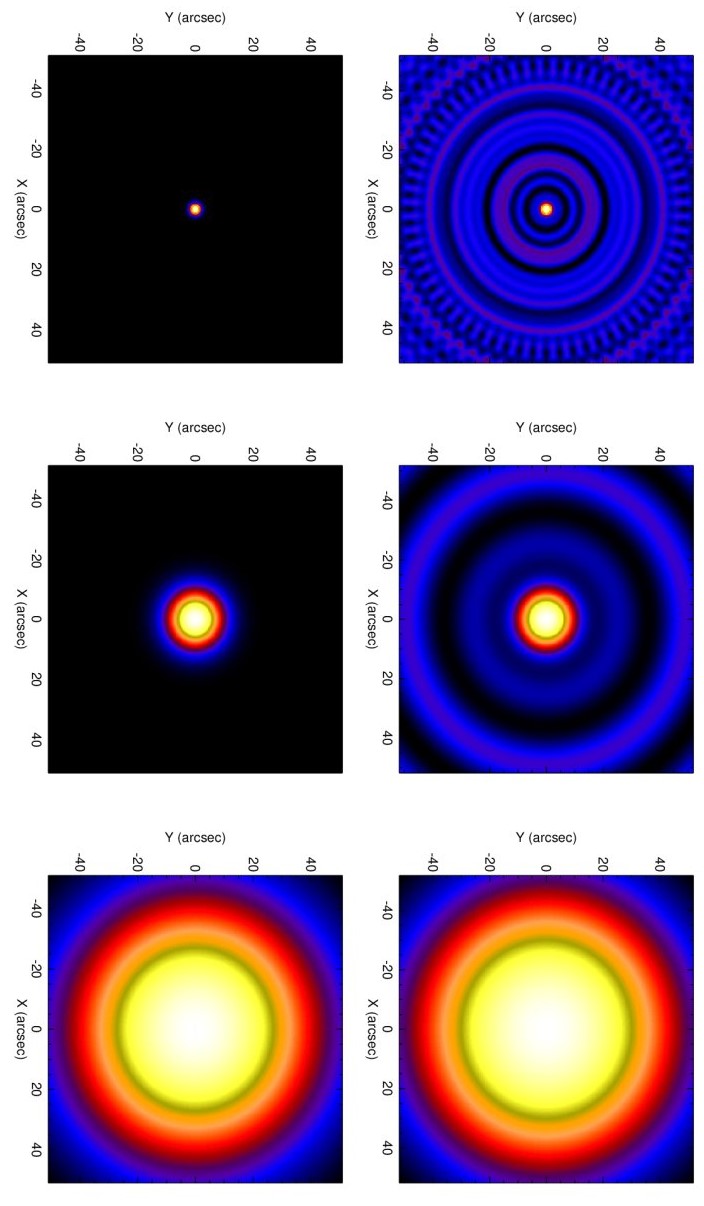} 
\caption{The computational tools of multi-scale CLEAN. Top row: the three PSF components, obtained by using Equation \eqref{psf-natural} with $N=3$ and according to the grouping of detectors as in Table \ref{table:parameters}. Bottom row: the three corresponding basis functions.}
\label{fig:PSFs-scale}
\end{center}
\end{figure}

\begin{figure}[h]
\begin{center}
\includegraphics[width=11.cm,angle=90]{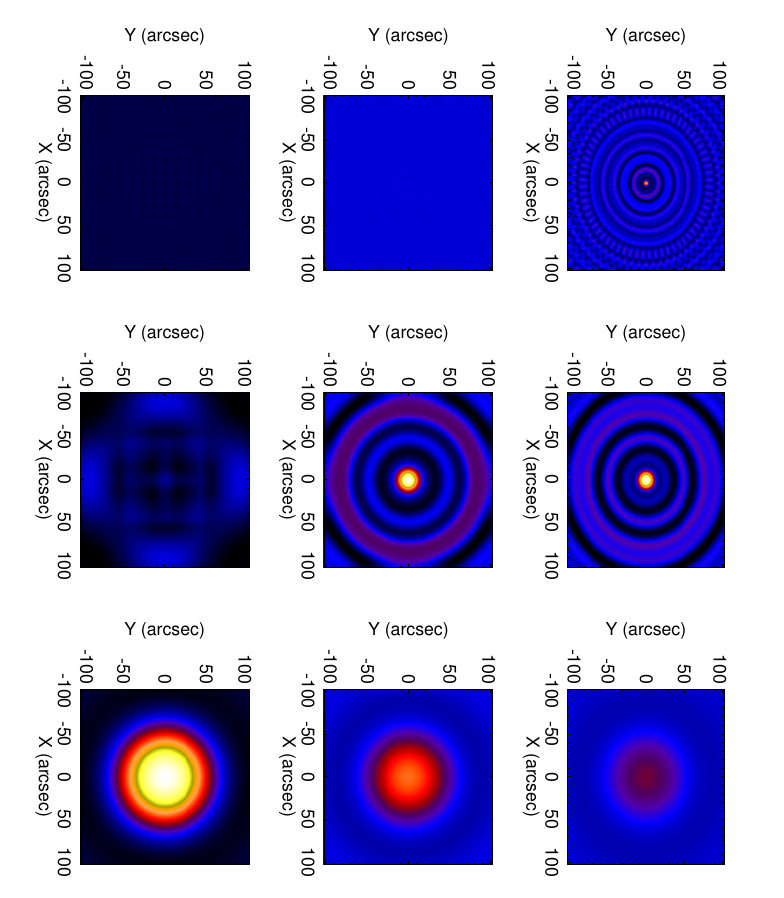} 
\caption{Cross-convolution products between PSF components and basis functions. In the panel, row $i$--column $j$ ($i,j=1,2,3$) contains cross-convolution $m_i*K_j$. In the first column ($j=1$) only the first image ($i=1$) is significantly different from zero, while in the second column ($j=2$) just the contribution of $i=3$ can be neglected.}
\label{fig:cross-conv}
\end{center}
\end{figure}

The validation of the multi-scale CLEAN reconstruction me\-thod performed in this section is concerned with the RHESSI imaging problem. This NASA instrument was operating until October 2018 and it is still the solar hard X-ray telescope with the best performances in terms of both spatial and spectral resolution. As said, the way its nine Rotating Modulation Collimators sampled the $(u,v)$ plane is described in Figure \ref{fig:uv-samples}, left panel, i.e. this telescope provided visibilities placed on nine circles with increasing radii in the spatial frequency plane. This implies that the RHESSI PSF can be written as the sum of up to nine PSF components, each one numerically computed as the inverse Fourier transform of the characteristic function of the set of $(u,v)$ points laying on each one of the sampled circles. As an example, in this paper the global PSF is decomposed as the sum of three PSF components (see Figure \ref{fig:PSFs-scale}, top row), the first one associated to the three circles with biggest radii, the second one associated to the three circles with intermedium radii, and the third one associated to the three circles with smallest radii. Correspondingly, we will consider three basis functions $m_1, m_2, m_3$ which are two-dimensional Gaussian functions whose standard deviations are determined by the utilized PSF components: specifically, we have computed the Full Width at Half Maximum (FWHM) of the central peak of each PSF component, we have exploited the relationship between FWHM and standard deviation for a Gaussian function ($FWHM=2\sqrt{2\ln 2} \sigma$), and used this standard deviation as the standard deviation of the corresponding scale (see Figure \ref{fig:PSFs-scale}, bottom row). All information characterizing this choice of the global PSF decomposition are contained in Table \ref{table:parameters}. Further, Figure \ref{fig:cross-conv} contains the nine cross-convolution products between the PSF components and the basis functions corresponding to the choices made in Figure \ref{fig:PSFs-scale}. As a numerical confirmation of Remark 3.3, the first and second colums of this figure contain two and one negligible cross-convolution products, respectively.

\begin{table}[h]
\begin{center}
\begin{tabular}{cccccc}
\hline\hline
 & detectors & $R_{max}$(arcsec$^{-1}$) & $\eta$ & resolution (arcsec) & FWHM (arcsec) \\
 \hline
 set 1 & $1,2,3$ & $0.221$ & $1$ & 2.26 & 2.8 \\
 \hline
 set 2 & $4,5,6$ & $0.0427$ & $0.19$ & 11.76 & 15.2\\
 \hline
 set 3 & $7,8,9$ & $0.029$ & $0.04$ & 61.08 & 76.8\\
 \hline\hline
\end{tabular}
\caption{Parameters describing the decomposition of the global RHESSI PSF for the experiment with synthetic data: 'detectors' denotes the circles composing each group of visibilities; $R_{max}$ is the radius of the circle with maximum radius in each group; $\eta$ is the scale bias factor; 'resolution' is the corresponding nominal spatial resolution as in \cite{hurford2003rhessi}, and FWHM is the Full Width at Half Maximum of the corresponding basis functions.}\label{table:parameters}
\end{center}
\end{table}


\subsection{Validation against simulated visibilities}
\begin{figure}[h]
\begin{center}
\includegraphics[width=10cm]{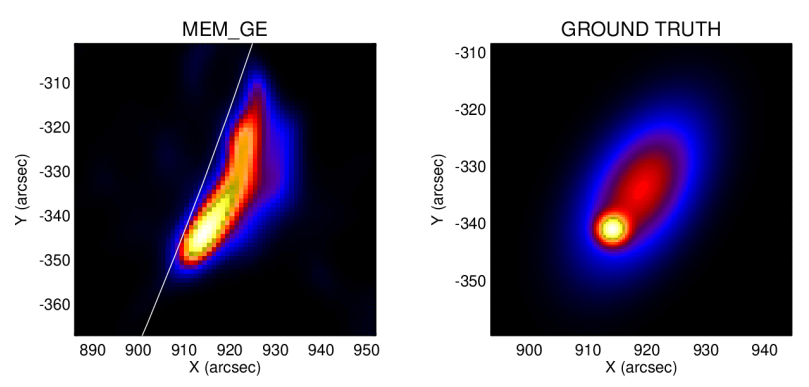} 
\caption{Validation of multi-scale CLEAN against simulated {\em{RHESSI}} data: the synthetic configuration, inspired by the December 2 2003 event in the time interval 22:54:00 - 22:58:00 UT. Left panel: the real source reconstructed by means of MEM$\_$GE \cite{massa2020mem_ge}. Right panel: the corresponding synthetic ground-truth used in the simulation experiment.}
\label{fig:simulation-1}
\end{center}
\end{figure}

In order to study the applicability conditions of our multi-scale CLEAN we performed a validation test based on the following two steps:
\begin{enumerate}
\item We invented a source configuration inspired by a specific real event observed by RHESSI, namely the one described in Figure \ref{fig:simulation-1}, occurred on December, 2 2003 in the time interval 22:54:00 - 22:58:00 UT, in the energy range 16-18 keV and showing an extended plus a compact source configuration. In the left panel of Figure \ref{fig:simulation-1}, this source is reconstructed from the RHESSI experimental visibilities by using a constrained maximum entropy algorithm \cite{massa2020mem_ge}. The right panel contains the synthetic source inspired by this reconstruction.
\item We then generated three synthetic 
data sets. Operationally, the Fourier transform has been applied three times from the flux distribution in the image plane to the set of visibilities sampled by RHESSI in the visibility space. Each time we considered a different level of statistics, i.e., a total flux per map equal to 1000, 10000, and 100000 photons cm$^{-2}$ sec$^{-1}$ arcsec$^{-2}$, respectively.
\item We randomly perturbed each visibility set $20$ times, by using a Gaussian distribution with constant standard deviation.
\end{enumerate}

\begin{figure}[h]
\begin{center}
\includegraphics[width=12.5cm]
{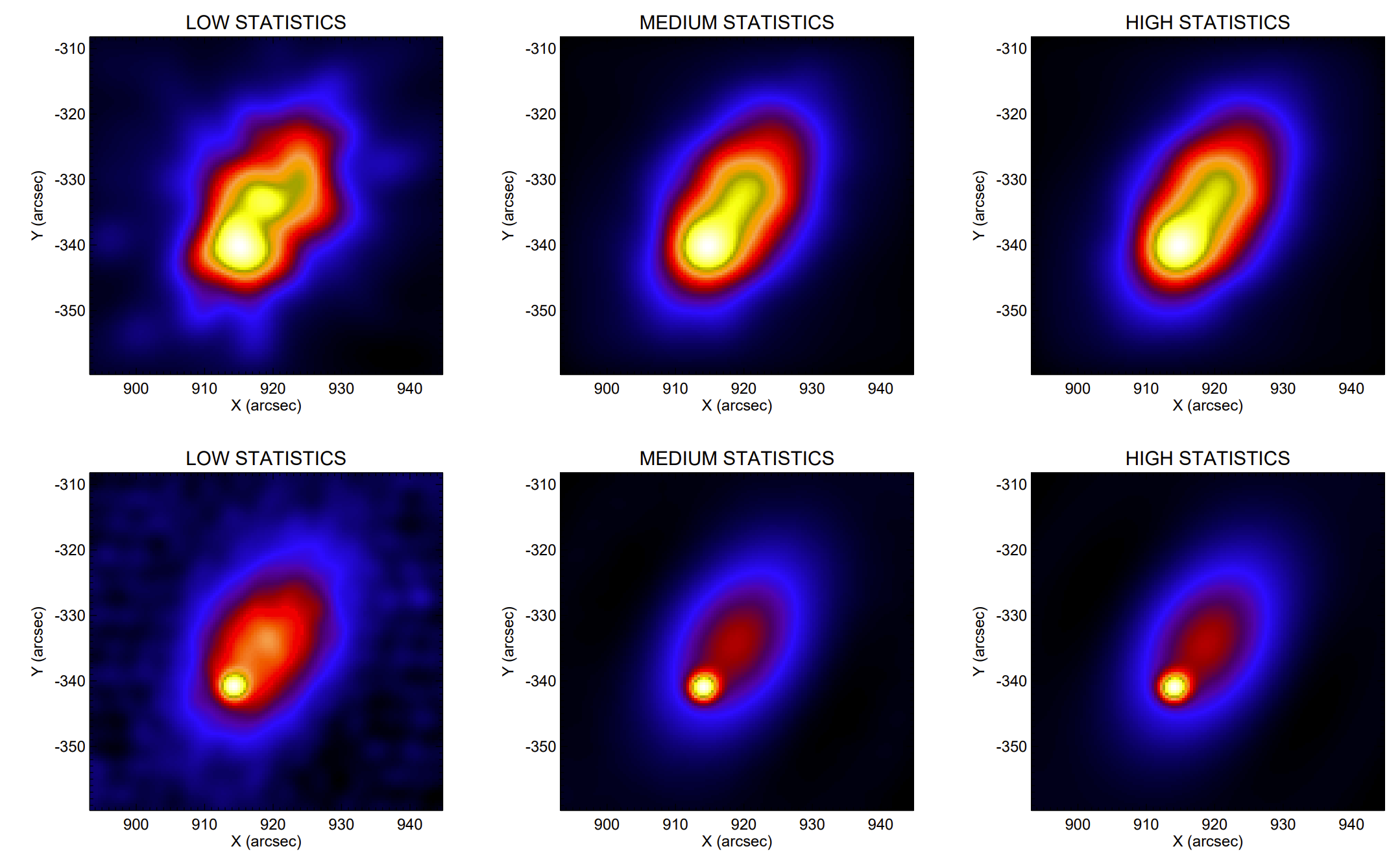} 
\caption{CLEAN vs multi-scale CLEAN for the reconstruction of the synthetic hard X-ray source represented in Figure \ref{fig:simulation-1}, right panel. Top row: reconstructions obtained by standard CLEAN in the case of low, medium and high statistics. Bottom row: reconstructions obtained by multi-scale CLEAN in the case of low, medium and high statistics. }
\label{fig:simulation-2}
\end{center}
\end{figure}

For each level of statistics, we applied the CLEAN and multi-scale CLEAN algorithms on the $20$ random realizations of the visibility set, and provided one representative reconstruction provided by each algorithm in Figure \ref{fig:simulation-2}. Further, Table \ref{multivsclean} contains the average values and standard deviations of the reconstructed imaging parameters. These results show, first, that multi-scale CLEAN is more robust than CLEAN with respect to the level of the noise affecting the input visibilities. Further, both algorithms perform well in identifying the location of both the extended and the compact source. On the other hand, multi-scale CLEAN outperforms standard CLEAN as far as photometry is concerned, since both the flux ratio and the peak ratio characterizing the reconstructions provided by the multi-scale approach are significantly closer to the ones characterizing the ground-truth. Finally, from an iconographical perspective, the reconstructions obtained by CLEAN seem to be significantly over-regularized, the ones provided by the multi-scale algorithm being able to better distinguish the compact source from the extended background halo.

\begin{table}[h]
\begin{center}
\resizebox{\columnwidth}{!}{
\begin{tabular}{cccccccc}
\hline\hline
& \multicolumn{2}{c}{First peak}  &  & \multicolumn{2}{c}{Second peak}  & Peaks ratio & Fluxes ratio\\
\cline{2-3}
\cline{5-6}
&  x & y & & x & y &  & \\
& (arcsec) & (arcsec) & & (arcsec) & (arcsec)  & & \\
\hline
\multicolumn{8}{c}{HIGH STATISTICS} \\
\hline

Ground truth &  914.0 & -341.0 & & 918.6 & -334.4 & 2.29 & 0.19\\
Multi-scale Clean &  914.0 $\pm$ 0.0 & -341.0 $\pm$ 0.0 & & 918.4 $\pm$ 0.0 & -334.6 $\pm$ 0.0 & 2.59 $\pm$ 0.01 & 0.18 $\pm$ 0.01\\
Clean        &  914.6 $\pm$ 0.0 & -340.2 $\pm$ 0.0 & & 915.2 $\pm$ 0.0 & -337.8 $\pm$ 0.0 & 1.15 $\pm$ 0.01 & 0.14 $\pm$ 0.00\\
\hline
\multicolumn{8}{c}{MEDIUM STATISTICS} \\
\hline
Ground truth & 914.0 & -341.0 & & 918.6 & -334.4 & 2.29 & 0.19\\
Multi-scale Clean & 914.0 $\pm$  0.0 & -341.0 $\pm$  0.0 & & 918.4 $\pm$ 0.0 & -334.8 $\pm$ 0.2 & 2.54 $\pm$ 0.04 & 0.18 $\pm$ 0.03\\
Clean        & 914.5 $\pm$ 0.2 & -340.3 $\pm$ 0.2 & & 915.2 $\pm$ 0.1 & -337.8 $\pm$ 0.0 & 1.14 $\pm$ 0.01& 0.14 $\pm$ 0.01\\
\hline
\multicolumn{8}{c}{LOW STATISTICS} \\
\hline
Ground truth & 914.0 & -341.0 & & 918.6 & -334.4 & 2.29 & 0.19\\
Multi-scale Clean  & 914.2 $\pm$ 0.4 & -340.9 $\pm$ 0.5 & & 916.5 $\pm$ 1.6 & -336.7 $\pm$ 1.5 & 1.60 $\pm$ 0.34 & 0.15 $\pm$ 0.04\\
Clean        & 914.7 $\pm$ 0.2 & -340.3 $\pm$ 0.3 & & 915.4 $ \pm$ 1.3 & -337.4 $ \pm$ 1.4 & 1.14 $ \pm$ 0.08 & 0.13 $ \pm$ 0.01 \\
\hline\hline
\end{tabular}}
\caption{The parameters reconstructed by CLEAN and multi-scale CLEAN are compared to the ones associated to the ground truth images. The compact source is referred to as first source, while the elongated source is referred to as second source.}\label{multivsclean}
\end{center}
\end{table}

\subsection{RHESSI experimental visibilities}

The standard model for solar flares \cite{tandberg1988physics,svestka2012solar} assumes that, at hard X-ray energies, a typical event is characterized by a low-energy extended source in the corona produced by thermal mechanisms, plus two high-energy compact sources down in the solar chromosphere produced by collisional bremsstrahlung \cite{warren2004thermal,prato2006regularized,saint2005thermal,hudson2011global}. However, on July 19 2012, RHESSI observed an M-class flare that, for the whole duration of the hard X-ray emission, was characterized by another coronal high-energy source over the thermal coronal loop-top. The analysis of this surprising observation has been performed in several papers \cite{oka2015electron,sun2014differential,gritsyk2016x,yuan2019compact,krucker2013particle,liu2013plasmoid}. In particular, the standard CLEAN algorithm used in \cite{liu2013plasmoid} was able to identify this high-energy coronal source, but not to clearly separate it from the close thermal emission. A better separation was obtained in \cite{krucker2013particle} by using standard CLEAN twice: first, using just the high-resolution detectors, and then using the coarser detectors after appropriate image subtraction. We applied our multi-scale CLEAN to RHESSI data corresponding to this event in two energy channels, ($6-8$ keV and $30-80$ keV). In this case we used all RHESSI detectors that were operational at that time (i.e., detectors from 3 through 9) according to two different grouping for the two energy channels: in the 6-8 keV channel, set 1 was made of detectors 3 and 4, set 2 of detectors 5 and 6, set 3 of detectors 7, 8, and 9; in the 30-80 keV channel, set 1 was made of detectors 3 and 4, set 2 of detectors 5, 6, and 7, set 3 of detectors 8 and 9. 

\begin{figure}
\includegraphics[width=12.7cm]{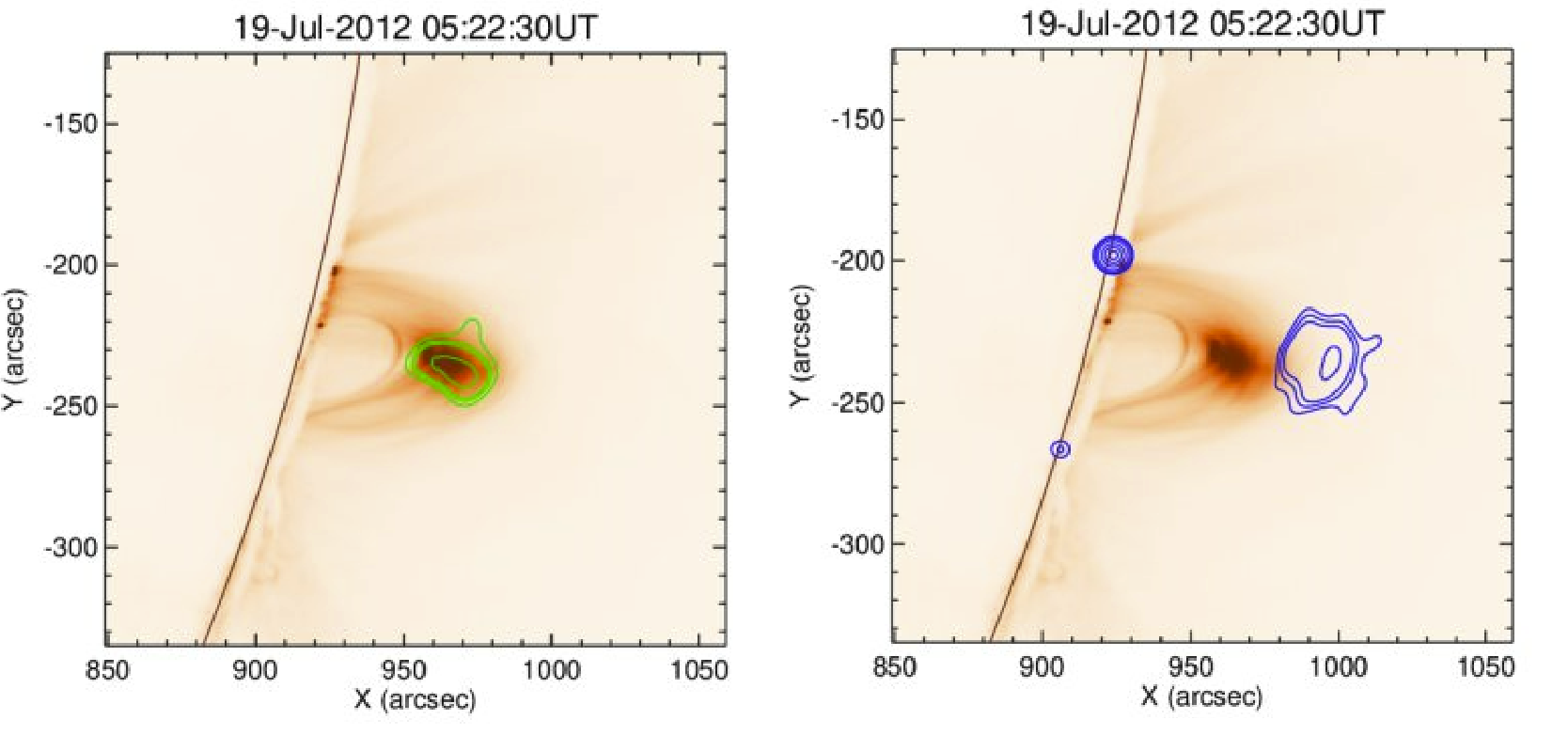}
\caption{The X-ray emission recorded by RHESSI on July, 19 2012 in the time interval from 05:21:00 to 05:24:00 UT. Left panel: the thermal coronal emission in the energy range between 6 and 8 keV (55, 65, 70, and 90$\%$ green contour levels). Right panel: the chromospheric footpoints and the high-energy coronal source in the energy range between 30 and 80 keV (15, 20, 30, 50, 70, and 90$\%$ blue contour levels for the foot-points; 2, 2.5, 3, and 5 blue contour levels for the coronal source). The hard X-ray emissions are superimposed to the 193 \AA{} extreme ultra-violet map recorded by the Atmospheric Imaging Assembly on-board the Solar Dynamics Obervatory (SDO/AIA).}\label{fig:july-av}
\end{figure}

Figure \ref{fig:july-av} shows the behavior of the multi-scale algorithm in the two energy ranges. In the low-energy case (left panel), where just one scale is present, the method is capable of adapting to this single scale, although it is set in such a way to seek for three different scales (in other terms, multi-scale CLEAN nicely behaves as CLEAN when it has to reconstruct one scale). In the high energy scale (right panel), the algorithm is perfectly able to distinguish the coronal source from the two footpoints in one step. Further, it nicely reproduces the asymmetry of the two chromospheric footpoints, which is due to the fact that part of the emission from the flare ribbon is occulted by the solar disk, while the spatial resolution obtained for these compact sources is better with respect to standard CLEAN. 

\section{Conclusions}
Hard X-ray imaging is a crucial Fourier-based modality for investigating the physics of high-energy emissions from the Sun and in the last decades several missions have been realized and conceived to launch these kind of telescopes in space. From the image processing perspective, a large amount of scientific literature in this field utilizes CLEAN to deconvolve the effects of the instrumental PSF. In this paper we introduce a genuinely multi-scale version of the iterative CLEAN algorithm, in which the input data are sets of photon visibilities (i.e. spatial Fourier components of the emission distribution), the cleaning of the image is performed in parallel for all the dirty maps components, the scale bias factor is determined by accounting for the peculiar way visibilities are recorded by the instrument, and the background is added to the final image in a very natural way.  We have validated the method against some sets of realistically simulated visibilities and proved its effectiveness in the case of experimental observation provided by RHESSI. 

The accuracy of this multi-scale CLEAN depends on the way the dirty map components are computed, i.e., on the way scales are identified and detectors are grouped. Therefore, it would be important to implement an iterative scheme for the automated identification of the optimal grouping of the detectors. Further, this general approach to multi-scale deconvolution can be probably extended to other deconvolution methods, like Pixon, that are frequently used in hard X-ray imaging, and to other imaging problems in which the global PSF can be modelled as the sum of a finite number of components.

\thanks{MP and AMM are and will always be grateful to Richard A. Schwartz who first inspired this research activity on multi-scale algorithms for hard X-ray imaging. Richard passed away in 2019 and this paper is devoted to his memory. MP and AMM also thanks Sara Giordano for her preliminary work on this topic made during her PhD period at the Dipartimento di Matematica, Università di Genova.}

 \bibliographystyle{unsrt}
\bibliography{mybiblio}

\begin{thebibliography}{10}

\bibitem{pantin2017deconvolution}
E.~Pantin, J.-L. Starck, and F.~Murtagh.
\newblock Deconvolution and blind deconvolution in astronomy.
\newblock {\em Blind Image Deconvolution}, pages 301--340, 2017.

\bibitem{bertero2021introduction}
M.~Bertero, P.~Boccacci, and C.~De~Mol.
\newblock {\em Introduction to inverse problems in imaging}.
\newblock CRC press, 2021.

\bibitem{kruger2012introduction}
A.~Kr{\"u}ger.
\newblock {\em Introduction to solar radio astronomy and radio physics},
  volume~16.
\newblock Springer Science \& Business Media, 2012.

\bibitem{piana2022hard}
M.~Piana, A.~G. Emslie, A.~M. Massone, and B.~R. Dennis.
\newblock {\em Hard X-Ray Imaging of Solar Flares}, volume 164.
\newblock Springer, 2022.

\bibitem{hogbom1974aperture}
J.~A. H{\"o}gbom.
\newblock Aperture synthesis with a non-regular distribution of interferometer
  baselines.
\newblock {\em Astronomy and Astrophysics Supplement, Vol. 15, p. 417}, 15:417,
  1974.

\bibitem{dennis2009hard}
B.~R. Dennis and R.~L. Pernak.
\newblock Hard x-ray flare source sizes measured with the ramaty high energy
  solar spectroscopic imager.
\newblock {\em The Astrophysical Journal}, 698(2):2131, 2009.

\bibitem{li2011application}
F.~Li, T.~J. Cornwell, and F.~de~Hoog.
\newblock The application of compressive sampling to radio astronomy-i.
  deconvolution.
\newblock {\em Astronomy \& Astrophysics}, 528:A31, 2011.

\bibitem{zhang2020parameterized}
L.~Zhang, L.~Xu, and M.~Zhang.
\newblock Parameterized clean deconvolution in radio synthesis imaging.
\newblock {\em Publications of the Astronomical Society of the Pacific},
  132(1010):041001, 2020.

\bibitem{chu2019deconvolution}
Z.~Chu, S.~Zhao, Y.~Yang, and Y.~Yang.
\newblock Deconvolution using clean-sc for acoustic source identification with
  spherical microphone arrays.
\newblock {\em Journal of Sound and Vibration}, 440:161--173, 2019.

\bibitem{bose2002sequence}
R.~Bose, A.~Freedman, and B.~D. Steinberg.
\newblock Sequence clean: A modified deconvolution technique for microwave
  images of contiguous targets.
\newblock {\em IEEE Transactions on Aerospace and Electronic Systems},
  38(1):89--97, 2002.

\bibitem{cornwell2008multiscale}
T.~J. Cornwell.
\newblock Multiscale clean deconvolution of radio synthesis images.
\newblock {\em IEEE Journal of selected topics in signal processing},
  2(5):793--801, 2008.

\bibitem{1992Sci...258..618A}
L.~{Acton}, S.~{Tsuneta}, Y.~{Ogawara}, R.~{Bentley}, M.~{Bruner},
  R.~{Canfield}, L.~{Culhane}, G.~{Doschek}, E.~{Hiei}, H.~{Hirayama},
  T.~Hudson, T.~{Kosugi}, J.~{Lang}, J.~{Lemen}, J.~{Nishimura},
  K.~{Makishima}, Y.~{Uchida}, and T.~{Watanabe}.
\newblock {The YOHKOH mission for high-energy solar physics}.
\newblock {\em Science}, 258(5082):618--625, October 1992.

\bibitem{ogawara1992status}
Y.~Ogawara, L.~W. Acton, R.~D. Bentley, M.~E. Bruner, J.~L. Culhane, E.~Hiei,
  T.~Hirayama, H.~S. Hudson, T.~Kosugi, J.~R. Lemen, et~al.
\newblock The status of yohkoh in orbit-an introduction to the initial
  scientific results.
\newblock {\em Publications of the Astronomical Society of Japan}, 44:L41--L44,
  1992.

\bibitem{lin2003reuven}
R.~P. Lin, B.~R. Dennis, G.~J. Hurford, D.~M. Smith, A.~Zehnder, P.~R. Harvey,
  D.~W. Curtis, D.~Pankow, P.~Turin, M.~Bester, et~al.
\newblock The reuven ramaty high-energy solar spectroscopic imager (rhessi).
\newblock {\em The Reuven Ramaty High-Energy Solar Spectroscopic Imager
  (RHESSI) Mission Description and Early Results}, pages 3--32, 2003.

\bibitem{fletcher2011observational}
L.~Fletcher, B.~R. Dennis, H.~S. Hudson, S.~Krucker, K.~Phillips, A.~Veronig,
  M.~Battaglia, L.~Bone, A.~Caspi, Q.~Chen, et~al.
\newblock An observational overview of solar flares.
\newblock {\em Space science reviews}, 159:19--106, 2011.

\bibitem{krucker2020spectrometer}
S.~Krucker, G.~J. Hurford, O.~Grimm, S.~K{\"o}gl, H.-P. Gr{\"o}belbauer,
  L.~Etesi, D.~Casadei, A.~Csillaghy, A.~O. Benz, N.~G. Arnold, et~al.
\newblock The spectrometer/telescope for imaging x-rays (stix).
\newblock {\em Astronomy \& Astrophysics}, 642:A15, 2020.

\bibitem{massa2022first}
P.~Massa, A.~F. Battaglia, A.~Volpara, H.~Collier, G.~J. Hurford, M.~Kuhar,
  E.~Perracchione, S.~Garbarino, A.~M. Massone, F.~Benvenuto, et~al.
\newblock First hard x-ray imaging results by solar orbiter stix.
\newblock {\em Solar Physics}, 297(7):93, 2022.

\bibitem{hurford2003rhessi}
G.~J. Hurford, E.~J. Schmahl, R.~A. Schwartz, A.~J. Conway, M.~J. Aschwanden,
  A.~Csillaghy, B.~R. Dennis, C.~Johns-Krull, S.~Krucker, R.~P. Lin, et~al.
\newblock The rhessi imaging concept.
\newblock {\em The Reuven Ramaty High-Energy Solar Spectroscopic Imager
  (RHESSI) Mission Description and Early Results}, pages 61--86, 2003.

\bibitem{massa2023stix}
P.~Massa, G.~J. Hurford, A.~Volpara, M.~Kuhar, A.~F. Battaglia, H.~Xiao,
  D.~Casadei, E.~Perracchione, S.~Garbarino, S.~Guastavino, et~al.
\newblock Stix imaging i--concept.
\newblock {\em arXiv preprint arXiv:2303.02485}, 2023.

\bibitem{kosugi1991hard}
T.~Kosugi, K.~Makishima, T.~Murakami, T.~Sakao, T.~Dotani, M.~Inda, K.~Kai,
  S.~Masuda, H.~Nakajima, Y.~Ogawara, et~al.
\newblock The hard x-ray telescope (hxt) for the solar-a mission.
\newblock {\em Solar Physics}, 136:17--36, 1991.

\bibitem{massa2020mem_ge}
P.~Massa, R.~A. Schwartz, A.~K. Tolbert, A.~M. Massone, B.~R. Dennis, M.~Piana,
  and F.~Benvenuto.
\newblock Mem\_ge: a new maximum entropy method for image reconstruction from
  solar x-ray visibilities.
\newblock {\em The Astrophysical Journal}, 894(1):46, 2020.

\bibitem{tandberg1988physics}
E.~Tandberg-Hanssen and A.~G. Emslie.
\newblock {\em The physics of solar flares}, volume~14.
\newblock Cambridge University Press, 1988.

\bibitem{svestka2012solar}
Z.~Svestka.
\newblock {\em Solar flares}, volume~8.
\newblock Springer Science \& Business Media, 2012.

\bibitem{warren2004thermal}
H.~P. Warren and S.~K. Antiochos.
\newblock Thermal and nonthermal emission in solar flares.
\newblock {\em The Astrophysical Journal}, 611(1):L49, 2004.

\bibitem{prato2006regularized}
M.~Prato, M.~Piana, J.~C. Brown, A.~G. Emslie, E.~P. Kontar, and A.~M. Massone.
\newblock Regularized reconstruction of the differential emission measure from
  solar flare hard x-ray spectra.
\newblock {\em Solar Physics}, 237(1):61--83, 2006.

\bibitem{saint2005thermal}
P.~Saint-Hilaire and A.~O. Benz.
\newblock Thermal and non-thermal energies of solar flares.
\newblock {\em Astronomy \& Astrophysics}, 435(2):743--752, 2005.

\bibitem{hudson2011global}
H.~S. Hudson.
\newblock Global properties of solar flares.
\newblock {\em Space Science Reviews}, 158(1):5--41, 2011.

\bibitem{oka2015electron}
M.~Oka, S.~Krucker, H.~S. Hudson, and P.~Saint-Hilaire.
\newblock Electron energy partition in the above-the-looptop solar hard x-ray
  sources.
\newblock {\em The Astrophysical Journal}, 799(2):129, 2015.

\bibitem{sun2014differential}
J.~Q. Sun, X.~Cheng, and M.~D. Ding.
\newblock Differential emission measure analysis of a limb solar flare on 2012
  july 19.
\newblock {\em The Astrophysical Journal}, 786(1):73, 2014.

\bibitem{gritsyk2016x}
P.~A. Gritsyk and B.~V. Somov.
\newblock X-ray and microwave emissions from the july 19, 2012 solar flare:
  Highly accurate observations and kinetic models.
\newblock {\em Astronomy Letters}, 42:531--543, 2016.

\bibitem{yuan2019compact}
D.~Yuan, S.~Feng, D.~Li, Z.~Ning, and B.~Tan.
\newblock A compact source for quasi-periodic pulsation in an m-class solar
  flare.
\newblock {\em The Astrophysical Journal Letters}, 886(2):L25, 2019.

\bibitem{krucker2013particle}
S.~Krucker and M.~Battaglia.
\newblock Particle densities within the acceleration region of a solar flare.
\newblock {\em The Astrophysical Journal}, 780(1):107, 2013.

\bibitem{liu2013plasmoid}
W.~Liu, Q.~Chen, and V.~Petrosian.
\newblock Plasmoid ejections and loop contractions in an eruptive m7. 7 solar
  flare: evidence of particle acceleration and heating in magnetic reconnection
  outflows.
\newblock {\em The Astrophysical Journal}, 767(2):168, 2013.

\end{thebibliography}

\end{document}